\documentclass{ws-procs9x6}
\newcommand{\be}{\begin{equation}}
\newcommand{\ee}{\end{equation}}
\newcommand{\ba}{\begin{eqnarray}}
\newcommand{\ea}{\end{eqnarray}}
\newcommand{\la}{\langle}
\newcommand{\ra}{\rangle}
\begin{document}

\title{Comparing extractions of Sivers functions}
\author{M.~Anselmino$^1$,
        M.~Boglione$^1$,
    J.~C.~Collins$^2$,
        U.~D'Alesio$^3$,
    A.~V.~Efremov$^4$,
    K.~Goeke$^5$,
        A.~Kotzinian$^1$,
    S.~Menzel$^5$,
    A.~Metz$^5$,\\          
    \hspace{-1.3cm}         
        F.~Murgia$^3$,
        A.~Prokudin$^1$,
    P.~Schweitzer$^5$,
    W.~Vogelsang$^{6,7}$,
    F.~Yuan$^7$\hspace{-1.5cm}$\phantom{l}$}
\address{
$^1$Universit\`a di Torino and INFN, Sezione di Torino, Italy \\
$^2$Penn State University, 104 Davey Lab, University Park PA 16802, U.S.A.\\
$^3$INFN, Sezione di Cagliari and Universit\`a di Cagliari, Italy\\
$^4$Joint Institute for Nuclear Research, Dubna, 141980 Russia\\
$^5$Institut f{\"u}r Theoretische Physik II, Ruhr-Universit{\"a}t Bochum,
    Germany\\
\hspace{-0.3cm}\
$^6$Physics Department, Brookhaven National Laboratory, Upton, NY 11973, U.S.A.\\
$^7$RIKEN BNL Research Center, Building 510A, BNL, Upton, NY 11973, U.S.A.}

\maketitle

\abstracts{
A comparison is given of the various recently published
extractions of the Sivers functions from the HERMES and COMPASS
data on single-transverse spin asymmetries in semi-inclusive deeply
inelastic scattering.}

%
%
\vspace{-11cm}\begin{flushright}
\begin{minipage}{0.5cm}{\tt 
	BNL-NT-05/42\\
 	RBRC-572\\
	RUB-TPII-14-05
}\end{minipage}
\end{flushright}\vspace{8cm}

\section{Introduction}
\label{Sec-1:introduction}

Single-spin asymmetries (SSA) in semi-inclusive deeply inelastic
scattering (SIDIS) off transversely polarized nucleon targets have been under intense
experimental investigation over the past few
years.\cite{HERMES-new}\cdash\cite{Avakian:2002td}
Substantial asymmetries have been reported in some cases, in particular,
with best statistics, by the HERMES collaboration for scattering off a
proton target.

The importance of 
SSA lies in the fact that they provide new insights into QCD and nucleon
structure.\cite{Sivers:1989cc}\cdash\cite{Belitsky:2002sm}
For instance, the asymmetry in SIDIS may
contain an angular dependence of the form $\sin(\phi-\phi_S)$, where
$\phi$ and $\phi_S$ denote respectively the azimuthal angles of the produced
hadron and the target polarization vector with respect to the axis defined by
the hard virtual photon.\cite{Boer:1997nt}
This angular dependence arises from the so-called Sivers effect\cite{Sivers:1989cc}
tightly related to notions of an intrinsic asymmetry in the parton transverse
momentum distribution and angular momenta.
Factorization theorems\cite{Collins:1981uk}\cdash\cite{Collins:2004nx}
proven to leading power in the photon virtuality $Q$
provide the basis for a QCD description of the process, and allow to
extract the Sivers function from SIDIS
data\cite{HERMES-new}\cdash\cite{Diefenthaler:2005gx}
and to use it for predictions for the SSA in the Drell-Yan (DY) process,
hopefully to be explored experimentally at RHIC, COMPASS and the GSI.
Comparisons of SIDIS and the DY process will be particularly important
for testing our understanding of the underlying physics, since it has
been predicted\cite{Collins:2002kn,Belitsky:2002sm} that the Sivers functions
appear with opposite signs in these two processes. The approach just outlined has
been followed recently in Refs.~[\refcite{Efremov:2004tp}--\refcite{in-preparation}].
In this note we compare the results of these
papers for the extracted Sivers functions
\be\label{Eq-02}
    \Delta^N f_{q/p^\uparrow}(x,{\bf p}_T^2)
    \equiv -\,\frac{2|{\bf p}_T|}{M_N}\,f_{1T}^{\perp a}(x,{\bf p}_T^2)
    \equiv -\,\frac{2|{\bf p}_T|}{M_N}\,q_T(x,{\bf p}_T^2)\;.
\ee

In the extractions of the Sivers functions from SIDIS
several simplifying approximations were common between the groups,
namely the neglect of the so-called ``soft factor''\cite{Ji:2004wu,Collins:2004nx}
and the Sivers antiquark functions.
Different approaches were, however,
followed in Refs.~[\refcite{Anselmino:2005nn}--\refcite{in-preparation}]
concerning the treatment of the dependence of the distributions
on transverse parton momenta.

The Sivers SSA is obtained\cite{Airapetian:2004tw,Bacchetta:2004jz} by
weighting the events entering the spin asymmetry with $\sin(\phi-\phi_S)$.
When analyzed in this way, however, specific models for the dependence on
parton transverse momenta need to be made in the theoretical expression.
By assuming that the transverse momentum dependence of the Sivers function is of
the form $f_{1T}^{\perp a}(x,{\bf p}_T^2)=f_{1T}^{\perp a}(x)\,G({\bf p}_T^2)$
and/or similarly for other distribution or fragmentation functions, the Sivers
SSA as defined at HERMES\cite{Airapetian:2004tw} can be written generically as
\be\label{Eq-01}
        A_{UT}^{\sin(\phi-\phi_S)} = (-2) \;
        \frac{\sum_a e_a^2\,x F_{\rm Siv}^a(x)\, D_1^{a/\pi}(z)}{
              \sum_a e_a^2\,x f_1^a(x)\,D_1^{a/\pi}(z)} \;.
\ee
The factor $(-2)$ is due to conventions\cite{Bacchetta:2004jz} and
$F_{\rm Siv}^q(x)$ is some functional depending on $f_{1T}^\perp$
and the model used for parton transverse momenta.

Notice that by including in addition a factor of $P_{h\perp}/M_N$ into
the weight in (\ref{Eq-01}) the resulting SSA can be interpreted
model-independently in terms of the transverse moment of the Sivers
function\cite{Boer:1997nt}
\be\label{Eq:Def-Siv-transv-mom}
        f_{1T}^{\perp(1)a}(x) \equiv \int\!{\rm d}^2{\bf p}_T\;
        \frac{{\bf p}_T^2}{2 M_N^2}\;f_{1T}^{\perp a}(x,{\bf p}_T^2)
        =
        - \int\!{\rm d}^2{\bf p}_T\;
        \frac{|{\bf p}_T|}{4M_N}\;\Delta^N f_{q/p^\uparrow}(x,{\bf p}_T^2)
        \;.
\ee
Such weighted SSA were argued to be less sensitive to Sudakov suppression
which can be important for predictions involving the Sivers function.
{\sl Preliminary} HERMES data for such SSA are available\cite{HERMES-new} and
were studied in Ref.~[\refcite{Efremov:2004tp}], where a first fit for the
transverse moment of the Sivers function (\ref{Eq:Def-Siv-transv-mom})
was obtained. The result of [\refcite{Efremov:2004tp}] is in good
agreement with the studies of SSA analyzed {\sl without} a power of $P_{h\perp}$
in the weight\cite{Airapetian:2004tw}\cdash\cite{Diefenthaler:2005gx}
reported
in Refs.~[\refcite{Anselmino:2005nn}--\refcite{in-preparation}]. The next Sections
review and compare the fit results for the Sivers functions extracted in the different
approaches in Refs.~[\refcite{Anselmino:2005nn}--\refcite{in-preparation}].

\newpage

\section{The approach of
Refs.~[\protect\refcite{Anselmino:2005nn},\protect\refcite{Anselmino:2005ea}]}
\label{Sec-2:Anselmino-et-al}

In Ref.~[\refcite{Anselmino:2005nn}]
the azimuthal angular dependence
(Cahn effect) of the SIDIS unpolarized cross section
was used to extract the widths of the Gaussian $p_T$-dependent parton
distribution (pdf) and fragmentation (ff) functions
respectively as $\langle p_T^2\rangle$ = 0.25 (GeV/$c$)$^2$ and
$\langle K_T^2\rangle$ = 0.2 (GeV/$c$)$^2$.
A first estimate of the Sivers functions was then obtained
by fitting the data on $ A_{UT}^{\sin(\phi-\phi_S)}$
observed by HERMES collaboration.\cite{HERMES-new,Airapetian:2004tw}
In Ref.~[\refcite{Anselmino:2005ea}] a novel fit on
the new HERMES data\cite{Diefenthaler:2005gx} together with data
from the COMPASS collaboration\cite{Alexakhin:2005iw} was performed.
In both fits the full exact kinematics was always adopted.
The Sivers function ($u$, $d$ quarks) was parameterized as:
\be
\Delta^N f_ {q/p^\uparrow}(x,{\bf p}_T^2)  =  2 \, {N}_q(x) \,
f_ {q/p} (x)\, g({\bf p}_T^2)\, h({\bf p}_T^2) \,  , \label{sivfac}
\ee
\be
N_q(x) =  N_q \, x^{a_q}(1-x)^{b_q} \,
\frac{(a_q+b_q)^{(a_q+b_q)}}{a_q^{a_q} b_q^{b_q}}\,, \hspace*{.2cm}
g({\bf p}_T^2) =
\frac{e^{-p_T^2/\langle p_T^2\rangle}}{\pi \langle p_T^2\rangle}\,.
\ee
Two options for the $h({\bf p}_T^2)$ function were considered, namely:
\be
\label{hpt}
(\text{a})\;\;h({\bf p}_T^2) = \frac{2p_T M_0}{p_T^2+M_0^2}\,,\;\;\;\;
(\text{b})\;\;h({\bf p}_T^2) = \sqrt{2e}\,\frac{p_T}{M'} e^{-p_T^2/M'^2}\,,
\ee
the latter allowing, at leading order in $p_T/Q$, to give for
$F_{\rm Siv}^a$ in Eq.~(\ref{Eq-01}):
\be\label{Eq:Fsiv-A}
        F_{\rm Siv}^a(x) = \frac{\sqrt{\pi}}{2}\,
        \frac{M_N}{ \sqrt{\langle \widehat {p_T^2} \rangle +
      \langle K^2_T\rangle/z^2}} \,f_{1T}^{\perp(1)a}(x)
        \;\;{\rm with}\;\; \langle\widehat {p_T^2}\rangle =
        \frac{\langle p_T^2\rangle}{1+\langle p_T^2\rangle/M'^2}\,.
\ee
In the fits, $f_{q/p}(x)$ was taken from the LO MRST01
set\cite{Martin:2002dr}, whereas Kretzer's set\cite{Kretzer:2000yf}
for the LO ff was used.
The 7 parameters  were then extracted as\cite{Anselmino:2005ea}:
\ba
N_{u} = 0.32 \pm  0.11 & a_{u} = 0.29 \pm  0.35 & b_{u} = 0.53 \pm
3.58 \nonumber \\
N_{d} = -1.0 \pm  0.12 & a_{d} = 1.16 \pm  0.47 & b_{d} = 3.77 \pm  2.59 \\
M'^2 =  0.55    \pm   0.38   \,  &
(M_0^2 =  0.32    \pm   0.25)  \,    & {\rm (GeV}/c)^2\,,
\nonumber
\ea
with a $\chi^2$ per degree of freedom ($\chi^2_{\rm dof}$) of 1.06.
The one-sigma band shown in Fig.~1 (Eq.~(\ref{hpt}b)) takes
into account the errors with their correlations.

These results were then used to give predictions for SSA measurable in
SIDIS and DY processes for various kinematical configurations.

These effects were also
invoked\cite{Anselmino:1994tv,Anselmino:1998yz,D'Alesio:2004up,Anselmino:2004ky}
to generate SSA for other processes in
hadron-hadron-collisions\cite{Adams:1991rw,Adams:2003fx}
although the status of factorization is less clear in this case.
Here we only point out that the SIDIS data are sensitive to much smaller $x$
values than the E704 (STAR) ones.

\section{The approach of Ref.~[\protect\refcite{Vogelsang:2005cs}]}
\label{Sec-3:Vogelsang+Yuan}

In Ref.~[\refcite{Vogelsang:2005cs}] it was assumed that the final hadron's
transverse momentum is entirely due to the transverse-momentum dependence
in the Sivers function. There is then no further assumption on the particular
form of this dependence; rather it is integrated out in order to
compare to the experimental data. The transverse momenta contributed by the other
factors in the factorization formula  will give some smearing effects which may
be viewed as ``sub-dominant''. (However, we emphasize that this will
not be true toward small $z$ where the
transverse momentum in the fragmentation functions will become
important, likely resulting in a suppression of the asymmetry at small $z$.)
The ``$1/2$-moments'' of the Sivers functions were then introduced
in Ref.~[\refcite{Vogelsang:2005cs}] in the fit to the experimental data:
\begin{equation}
\label{Eq:VY-00}
q_T^{(1/2)}(x)\equiv \int d^2{\bf p_T}\frac{|{\bf p_T}|}{M_N}
f_{1T}^{\perp q}(x,{\bf p}_T^2) \ .
\end{equation}
These appear in an expression of the form (\ref{Eq-01}) for the
Sivers asymmetry, where
\be\label{Eq:VY-01}
        F_{\rm Siv}^q(x) = \frac{1}{2}\; q_{T}^{(1/2)}(x)\ .
\ee

In the actual fit to the HERMES data in\cite{Vogelsang:2005cs}
the functions $q_T^{(1/2)}(x)$ were modeled in terms of the unpolarized
$u$-quark distribution as
\be
\frac{u_T^{(1/2)}(x)}{u(x)}=S_u x(1-x)\ ,\hspace*{5mm}
\frac{d_T^{(1/2)}(x)}{u(x)}=S_d x(1-x)\ ,\label{eq9}
\ee
where $u(x)$ was taken from the GRV LO parameterizations for the unpolarized
parton distributions.\cite{GRV94} Furthermore, Kretzer's set for the LO
fragmentation functions\cite{Kretzer:2000yf} was used.
The fit to the new preliminary HERMES data gave
\begin{equation}
S_u=-0.81\pm 0.07,~~~S_d=1.86\pm 0.28 \ ,\label{eq19}
\end{equation}
with $\chi^2_{\rm dof}\approx 1.2$. A fit to the old published HERMES
data gave instead $S_u=-0.55\pm 0.37$ and $S_d=1.1\pm 1.6$, with a
similar size of $\chi^2_{\rm dof}$. The COMPASS data were not
included in the fit performed in\cite{Vogelsang:2005cs}, but a
comparison of the fit with the data was given, showing good agreement.
The results of the fit to the HERMES data were furthermore
used for making predictions for the SSAs in the Drell-Yan process
and in di-jet and jet-photon correlations at RHIC.

\newpage

\section{The approach of Refs.~[\protect\refcite{Collins:2005ie},\protect\refcite{in-preparation}]}
\label{Sec-4:Collins-et-al}

In Ref.~[\refcite{Collins:2005ie,in-preparation}] the distributions of
transverse parton momenta in $f_1^a$, $f_{1T}^{\perp a}$ and $D_1^a$
were assumed to be Gaussian with the respective widths
$\langle p_T^2\rangle$,           
$\langle p_T^2\rangle_{\rm Siv}$  
and $\langle K_T^2\rangle$        
taken to be flavour- and $x$- or $z$-independent.
In this model the $F_{\rm Siv}^a$ defined in (\ref{Eq-01}) is given by
the expression in Eq.~(\ref{Eq:Fsiv-A}) with $\langle\widehat{p_T^2}\rangle$
replaced by $\langle p_T^2\rangle_{\rm Siv}$.

The values $\langle K_T^2\rangle=0.16\,({\rm GeV}/c)^2$,
$\langle p_T^2\rangle=0.33\,({\rm GeV}/c)^2$ 
were extracted\cite{Collins:2005ie} from the HERMES data\cite{Airapetian:2002mf}
on $\la P_{h\perp}\ra$ and are similar to those discussed in
Sec.~\ref{Sec-2:Anselmino-et-al},
while $\langle p_T^2\rangle_{\rm Siv}\in[0.01;0.32]\,({\rm GeV}/c)^2$ remained
poorly constrained by positivity\cite{Bacchetta:1999kz} -- still allowing
an extraction of the {\sl transverse moment} of the Sivers function
(\ref{Eq:Def-Siv-transv-mom}).

In order to reduce the number of fit parameters the
prediction\cite{Pobylitsa:2003ty}
from the limit of a large number of colours $N_c$ was imposed:
\be\label{Eq:CE-02}
          f_{1T}^{\perp u}(x,{\bf p}_T^2) =
        - f_{1T}^{\perp d}(x,{\bf p}_T^2) \;\;\;
        \mbox{modulo $1/N_c$ corrections.}\ee
The best fit\cite{Collins:2005ie}
(using parameterizations\cite{Gluck:1998xa,Kretzer:2001pz})
to the {\sl published data}\cite{Airapetian:2004tw} is
\be\label{Eq:CE-03}
    xf_{1T}^{\perp(1)u}(x)
    \stackrel{\rm ansatz}{=}Ax^b(1-x)^5
    \stackrel{\rm fit}{=}-0.17x^{0.66}(1-x)^5
\ee
with a $\chi^2_{\rm dof}\sim 0.3$, and a 1-$\sigma$ uncertainty of roughly $\pm30\%$.
This result agrees well with the fit to the preliminary $P_{h\perp}$-weighted HERMES
data\cite{HERMES-new}, which were analyzed in a (transverse parton momentum)
model-independent way [\refcite{Efremov:2004tp}]. The good agreement of the results
in Refs.~[\refcite{Efremov:2004tp,Collins:2005ie}] is an important cross check
for the applicability of the Gauss model to the description of SSA in SIDIS.

For sake of a better comparison to the results by the other
groups\cite{Anselmino:2005nn,Anselmino:2005ea,Vogelsang:2005cs} the above fit
procedure was applied\cite{in-preparation} to the most recent and more precise
{\sl preliminary} HERMES data.\cite{Diefenthaler:2005gx}
The new fit has a $\chi^2_{\rm dof}\sim 2$ and is consistent\cite{in-preparation}
with that quoted in Eq.~(\ref{Eq:CE-03}).
One has to keep in mind that the large-$N_c$ relation (\ref{Eq:CE-02})
is a useful constraint at the {\sl present stage}, and will have to be
relaxed when future more precise data will become available.

Note that for $\langle K_T^2\rangle\to 0$ in (\ref{Eq:Fsiv-A}) one obtains
$F_{\rm Siv}^a(x)\to\frac12f_{1T}^{\perp(1/2)a}(x)$ within the Gaussian model.
This limit means that the produced hadron acquires no additional transverse
momentum from the fragmentation process, i.e.\
$D_1^a(z,{\bf K}_T^2)=D_1^a(z)\,\delta^{(2)}({\bf K}_T)$.
In this sense, the approach of Ref.~[\refcite{Vogelsang:2005cs}] discussed
in Sec.~\ref{Sec-3:Vogelsang+Yuan}, c.f. Eq.~(\ref{Eq:VY-01}), is contained
as a limiting case in the Gauss ansatz.

\newpage

\begin{figure}[ht]\epsfxsize=10cm
\centerline{\hspace{-0.2cm}
    \epsfxsize=2.5in\epsfbox{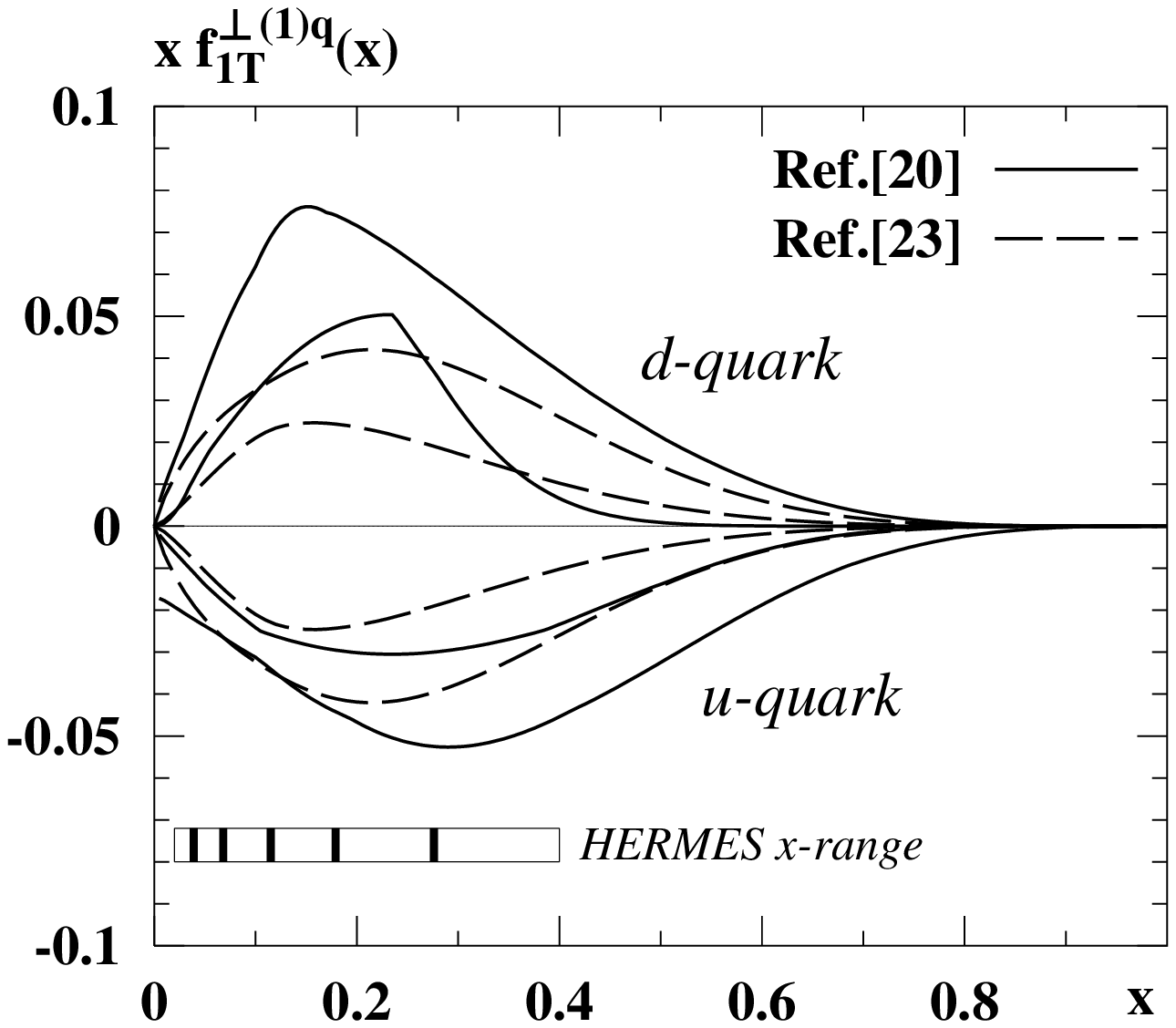}\hspace{-0.6cm}
    \epsfxsize=2.5in\epsfbox{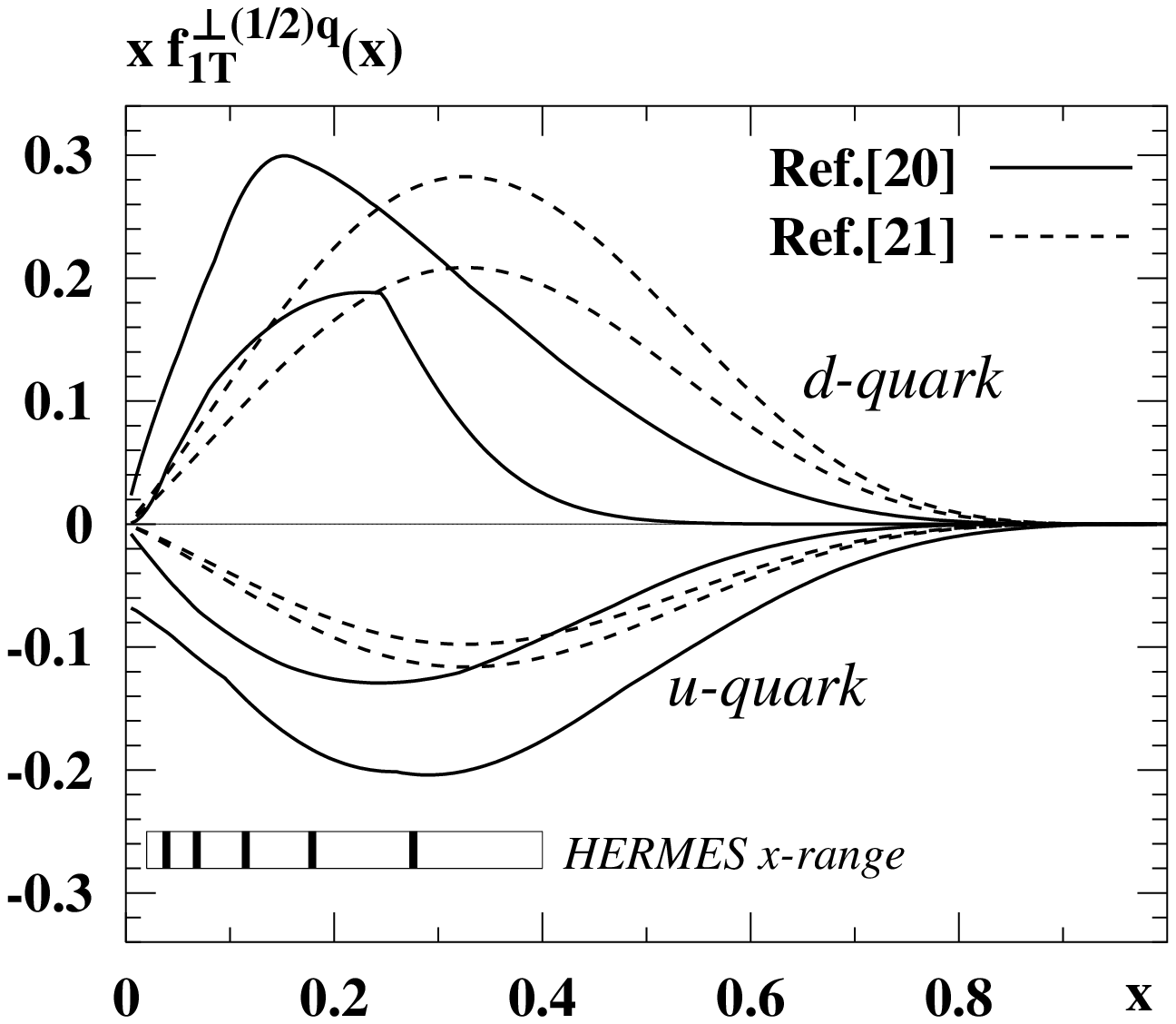}}
\caption{\label{Fig3-CE}
    The first and 1/2-transverse moments of the Sivers quark distribution functions,
    defined in Eqs.~(\ref{Eq:Def-Siv-transv-mom},~\ref{Eq:VY-00}), as extracted in
    Refs.~[\protect\refcite{Anselmino:2005ea,Vogelsang:2005cs,in-preparation}].
    The fits were constrained mainly (or solely) by the preliminary HERMES
    data\protect\cite{Diefenthaler:2005gx} in the indicated $x$-range.
    The curves indicate the 1-$\sigma$ regions of the various parameterizations.}
\end{figure}

\section{Comparison of the results and Conclusions}

It should be stressed that the various fit results, when used within the respective
approaches, provide equally good descriptions of the HERMES and COMPASS data. Here
we compare only those analyses\cite{Anselmino:2005ea,Vogelsang:2005cs,in-preparation}
in which the most recent and more precise {\sl preliminary} HERMES
data\cite{Diefenthaler:2005gx} were used.

In Fig.1a we compare the fits for $f_{1T}^{\perp(1)q}$ from
Refs.~[\refcite{Anselmino:2005ea,in-preparation}], and
in Fig.1b the fits for $f_{1T}^{\perp(1/2)q}$ from
Refs.~[\refcite{Anselmino:2005ea,Vogelsang:2005cs}].
(A direct comparison of [\refcite{Vogelsang:2005cs}] and
[\refcite{in-preparation}] is not possible.)
In view of the different models assumed for the transverse parton momenta and
the varying fit Ans\"atze, we observe a satisfactory {\sl qualitative} agreement ---
in the $x$-region constrained by the HERMES data.
However, a closer look reveals differences between the results in Fig.~1,
which indicate the size of the systematic uncertainties of
the three Sivers function fits mainly due to the use of different models
for the parton transverse momenta. These uncertainties were not estimated
in Refs.~[\refcite{Anselmino:2005ea,Vogelsang:2005cs,in-preparation}].

We have presented a comparison of three
extractions\cite{Anselmino:2005ea,Vogelsang:2005cs,in-preparation}
of Sivers functions from HERMES and COMPASS data
on single-transverse spin asymmetries in SIDIS. The three approaches
somewhat differ, but they describe the data with similar quality.
The fits are in good qualitative agreement, though there are differences
with regard to the size and shape of the extracted Sivers functions.
These differences reflect the model dependence of the fit results
which gives rise to a certain theoretical systematic uncertainty of the fit results.
The latter 
seems, however, less dominant than the statistical uncertainty of the fits
at the present stage.

It is clear that further information from experiment
will be vital. For now, one cannot really expect to obtain
much more than a first qualitative picture of the Sivers
functions. We also emphasize that it will be crucial for
the future to experimentally confirm the leading-power nature
of the observed spin asymmetries. For this, forthcoming COMPASS
or JLab data for scattering off a proton target and studies of the
$Q^2$-dependence of the asymmetries will be important.

The good qualitative agreement between the different approaches
observed here means that the
predictions\cite{Efremov:2004tp}\cdash\cite{in-preparation} for
the magnitude of the Sivers effect in DY are robust --- in the
kinematic region constrained by the HERMES data. This solidifies
the conclusions\cite{Efremov:2004tp}\cdash\cite{in-preparation}
that the predicted sign reversal of the Sivers function between
SIDIS and DY, can be tested in running or future experiments at
RHIC, COMPASS and PAX.

\paragraph*{Acknowledgments.}
U.D. and F.M. acknowledge partial support by MIUR under Cofinanziamento PRIN 2003.
A.E. is supported by grants RFBR 03-02-16816 and DFG-RFBR 03-02-04022.
J.C.C. is supported in part by the U.S. D.O.E., and by a Mercator
Professorship of DFG. W.V. and F.Y. are grateful to RIKEN,
Brookhaven National Laboratory and the U.S. Department of Energy
(contract number DE-AC02-98CH10886) for providing the facilities
essential for the completion of their work.
The work is partially supported by the European Integrated Infrastructure Initiative
Hadron Physics project under contract number RII3-CT-2004-506078.

\end{document}